\documentclass[aps,pra,a4paper,reprint,showpacs,nofootinbib,floatfix,floats]{revtex4-1}
\usepackage{pifont}
\usepackage{color}
\usepackage{amsmath}
\usepackage{amssymb}
\usepackage{amsfonts}
\usepackage{times,txfonts}
\usepackage{bbold}
\usepackage{hyperref}
\hypersetup{
    colorlinks=true,
    citecolor=blue,
    linkcolor=blue,
    urlcolor=blue
}
\usepackage{color}
\usepackage{bm}
\usepackage{graphicx}
\usepackage{float}
\usepackage{epsfig}
\usepackage{verbatim}
\usepackage{epstopdf}
\usepackage{natbib}
\usepackage{appendix}

\providecommand{\openone}{\leavevmode\hbox{\small1\kern-3.8pt\normalsize1}}

\def\ket#1{|#1 \rangle}

\begin{document}
\title{Two different types of optical hybrid qubits for teleportation in a lossy environment
}
\author{Hoyong Kim, Seung-Woo Lee, and Hyunseok Jeong}

\affiliation{Center for Macroscopic Quantum Control, Department of Physics and Astronomy, Seoul National University, Seoul 151-742, Korea}

\date{\today}

\begin{abstract}
We investigate the performance of quantum teleportation under a lossy environment using two different types of optical hybrid qubits. One is the hybrid of the vacuum and single-photon states with coherent states and the other is the hybrid of polarized single-photon states with coherent states. We have shown that the hybrid qubit using vacuum and single-photon states is generally  more robust to photon loss effects compared to the one using the photon polarization with respect to fidelities and success probabilities of quantum teleportation. 
\end{abstract}


\maketitle

\section{Introduction}
\label{sec:introduction}

Quantum teleportation is a protocol to transfer an unknown qubit from one place to another via an entangled quantum channel \cite{BennettPRL1993,exp1998}.
It is at the heart of various applications in quantum communication and computation. In particular, it plays a crucial role in implementing all-optical quantum computation \cite{Knill2001,Gottesman1999,Dawson2006,Hayes2010,Lund2008,Ralph2010}.
A typical qubit for optical quantum teleportation utilizes the horizontal and vertical polarization states of a single photon, \{$|H\rangle$, $|V\rangle$\} \cite{exp1998,Knill2001,Dawson2006,Hayes2010}, or alternatively the vacuum and single-photon states, \{$|0\rangle$, $|1\rangle$\} \cite{Lee00,Lund02}.
However, in this type of approaches based on a single-photon qubit, the success probability of a Bell-state measurement, which is an essential element in realizing the quantum teleportation protocol, cannot exceed 1/2 using linear optics and photon detection \cite{Calsa2001,Lutkenhaus99}.
Efforts are being made to overcome this limitation using auxiliary states, additional operations or multipartite encoding \cite{Grice2011,Zaidi2013,Ewert2014,Lee15} while each of them has its own price to pay.
 An alternative approach employs coherent states as the qubit basis, \{$\ket{\alpha}$, $\ket{-\alpha}$\} \cite{vanEnkPRA2001,JeongPRA2001}, where $\pm\alpha$ are amplitudes of the coherent states. It enables one to implement a nearly deterministic Bell-state measurement  \cite{JeongPRA2001,Jeong2002QIC,Ralph03}. However, due to the non-orthogonality of two coherent states, $|\alpha\rangle$ and $|-\alpha\rangle$, a necessary operation to finish the teleportation process such as the Pauli-Z operation cannot be performed in a deterministic way and produces additional errors \cite{Lund2008,Ralph2010}.

Recently, a hybrid approach to optical quantum information processing was proposed by combining advantages of the two aforementioned approaches \cite{Lee13}. In this approach, the logical qubit is constructed using entanglement between the polarization states of a single photon and coherent states that leads to nearly deterministic quantum controls \cite{Lee13}. 
It enables one to perform a near-deterministic quantum teleportation as well as near-deterministic universal gate operations in a more efficient manner compared to previous approaches \cite{Knill2001,Gottesman1999,Dawson2006,Hayes2010,Lund2008,Ralph2010}. 
The required resource is hybrid states in the form of
$|{\rm H_I}\rangle=(|H\rangle|\alpha\rangle+|V\rangle|-\alpha\rangle)/\sqrt{2}$  \cite{Lee13}.
Within this context, it was shown that such a hybrid entanglement is useful for teleportation between a polarized single-photon qubit and a coherent state qubit \cite{Park12} and for a loophole free Bell inequality test \cite{Kwon13}. 
However,  it is known that the generation of entanglement between a polarized single photon and coherent states
such as $|{\rm H_I}\rangle$ is highly demanding \cite{Gerry1999,Jeong2005,Munro2005,Shapiro2007}.
There exists a recent theoretical proposal that enables one to efficiently generate the state  $|{\rm H_I}\rangle$ based on parametric downconversion, linear optics elements, and photodetectors  \cite{Kwon15} while it requires preparation of a coherent-state superposition \cite{Ourj} as a resource.
On the other hand, the hybrid entanglement of the vacuum and single photon (instead of single-photon polarization) with coherent states, such as $|{\rm H_{II}}\rangle=(|0\rangle|\alpha\rangle+|1\rangle|-\alpha\rangle)/\sqrt{2}$, was successfully demonstrated in recent experiments \cite{Jeong14,Morin14}.
In addition, a previous study showed that the qubits utilizing the vacuum and single-photon states are more robust against losses compared to the polarized single-photon qubits  \cite{hoyong14}.
We therefore need to investigate whether the approach based on the hybrid state in the form of $|{\rm H_{II}}\rangle$ is equivalently useful, or even more useful, compared to the one using state $|{\rm H_I}\rangle$ for quantum information processing.

In this paper, in order to compare these two different types of hybrid qubits, we  consider implementations of quantum teleportation in a lossy environment. We analyze the environmental effects caused by photon losses on the entangled channel distributed between two separated parties. 
Our analysis shows that the quantum teleportation with the hybrid of vacuum and single photon with coherent state
is more robust to photon losses than the hybrid of photon polarization with coherent states.

\section{Two types of optical hybrid qubits}
\label{sec:hybridqubits}

Since there are a number of studies on quantum information processing using various kinds of optical hybrid systems \cite{Lee13,Jeong14,Morin14,Martini98,Martini08,Sekatski10,Sekatski12,Ghobadi13,Bruno13,Lvovsky13,Andersen13,Sheng13,furusawa1,furusawa2}, we first need to clarify the types of optical hybrid qubits that we consider in this paper. 
The first one is the hybrid of photon polarization with coherent states, which was originally used to propose the hybrid scheme of optical quantum information processing recently \cite{Lee13}. The other is the hybrid of vacuum and single photon with coherent states, which was recently generated by experiments \cite{Jeong14,Morin14}.
We consider optical hybrid qubits constructed in the logical basis, 
\begin{align}
\label{eq:qbasis}
  \{|0_L\rangle=|+\rangle|\alpha\rangle, |1_L\rangle=|-\rangle|-\alpha\rangle\},
\end{align}
and the two different types of hybrid qubits are then defined as
\begin{itemize}
\item[I.] the hybrid of the single-photon polarization with coherent states where $|\pm\rangle=(|H\rangle\pm|V\rangle)/\sqrt{2}$,
\item[II.] the hybrid of the vacuum and single photon with coherent states where $|\pm\rangle=(|0\rangle\pm|1\rangle)/\sqrt{2}$.
\end{itemize}
We will refer to the former as {\em type-I} hybrid qubit which is the same form used in Ref.~\cite{Lee13}, while the latter will be referred to as {\em type-II} hybrid qubit hereafter.

It was shown that using type-I hybrid qubits, quantum teleportation and a universal set of quantum computation can be implemented in a nearly deterministic way using passive linear optics elements and photodetectors \cite{Lee13}.
This scheme was shown to outperform previous all-optical schemes \cite{Dawson2006,Hayes2010,Lund2008,Ralph2010}
when considering together the resource requirements and the fault tolerance limits with photon losses.
In principle, the same structure of quantum teleportation and computation schemes can be constructed using type-II qubits.
In the type-II qubit based approach for quantum teleportation, however, there are a couple of points to note.
In order to complete the quantum teleportation process, the Pauli X and Pauli Z operations are required.
The Pauli X operation for a type-II hybrid qubit in the logical basis (\ref{eq:qbasis}) can be implemented deterministically by acting the $\pi$-phase shift on each of the two modes.
On the other hand, a flip between $|0\rangle$ and  $|1\rangle$  (i.e., $|0\rangle \leftrightarrow |1\rangle$) is required to perform the Pauli-Z operation;
this cannot be performed deterministically using linear optics elements.
One simple working solution, for the moment, is to ``logically relabel'' the vacuum and the single photon, $|0\rangle$ and $|1\rangle$, whenever it is necessary. In other words, we know that $|0\rangle$ and $|1\rangle$ remain unaltered, whenever they should be altered, so that it can be logically corrected at the final measurement stage.
Beside this point, fortunately, if we take the logical qubit basis as the form in Eq.~(\ref{eq:qbasis}), it is possible to perform the hybrid qubit teleportation with the same success probability with the type-I hybrid qubit \cite{Lee13}, when there is no loss, and with better success probabilities when photon losses occur. 
In the following sections, we will analyze the teleportation protocols and effects of lossy environments for two different types of hybrid qubits in further detail.

\section{Quantum teleportation for hybrid qubits under photon losses}
\label{sec:teleportation}

In the standard quantum teleportation procedure \cite{BennettPRL1993}, Alice is supposed to teleport an arbitrary unknown state $|\phi\rangle=\mu|0_L\rangle+\nu|1_L\rangle$ to Bob via  a maximally entangled quantum channel $|\Psi_{ch}\rangle=(|0_L\rangle|0_L\rangle+|1_L\rangle|1_L\rangle)/\sqrt{2}$. Alice performs a Bell-state measurement on the unknown qubit and her part of the entangled channel, and sends the measurement outcome to Bob. Bob applies an appropriate unitary transform on his state depending on Alice's measurement outcome in order to reconstruct the original qubit.
In an ideal situation, quantum teleportation can be carried out  with the unit success probability and the teleported state should be exactly the same to the input state. However, in realistic implementations, there are factors that reduce  the success probability and the teleportation fidelity. Here, we consider two major such factors. One is  inefficiency of the Bell-state measurement and the other is photon losses in the quantum channel.
In the following subsections, we will calculate and compare the fidelities between the input and the output states and the success probabilities of teleportation for two different types of hybrid qubits.

\begin{figure*}[t]
		\epsfig{file=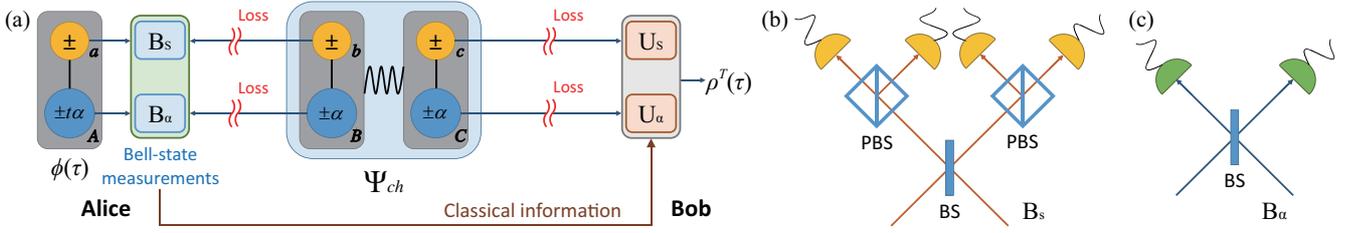,width=17.5cm}
		\caption{(Color online) (a) Schematic of hybrid quantum teleportation. We use the dynamical orthonormal basis for the input state $|\phi(\tau)\rangle$ in order to reflect the amplitude damping of coherent states of the channel state. ${\rm U_s}$ (${\rm U_\alpha}$) represents a unitary transform applied on a single-photon state (coherent state) and $\rho^T(\tau)$ represents the final teleported state. (b) The Bell-state measurement of photon-polarized states, $\rm B_s$, for type-I hybrid qubits can be performed using the 50:50 beam splitter (BS), two polarizing beam splitters (PBS), and four photodetectors. (c) The Bell-state measurement for the coherent states, $\rm B_\alpha$, can be implemented by the 50:50 beam splitter and two photon number resolving detectors. Another Bell-state measurement of vacuum and single-photon states, $\rm B_s$, for type-II hybrid qubits can be done as $\rm B_\alpha$ with the 50:50 beam splitter and two single-photon detectors.}
		\label{scheme}
\end{figure*}

\subsection{Teleportation of type-I hybrid qubits}
\label{sec:teleportation1}
Before considering photon loss effects for teleportation with type-I qubits, 
we briefly review the hybrid  teleportation scheme without loss described in Ref.~\cite{Lee13}. 
As shown in Fig.~\ref{scheme}, Alice and Bob share a hybrid entangled channel in order to teleport a type-I hybrid qubit from Alice to Bob. 
If there is no loss, the total product state of an unknown input state $|\phi\rangle$ and the channel state $|\Psi_{ch}\rangle$ in terms of the type-I hybrid encoding can be rewritten as 
\begin{align}
\label{eq:prod}
\nonumber
|\phi\rangle_{aA}|\Psi_{ch}\rangle_{bBcC} =&\frac{1}{4}\Bigg[\Bigg(\frac{|\Phi_{\rm P}^+\rangle_{ab}|\Phi_{\rm C}^+\rangle_{AB}}{N^+_\alpha}+\frac{|\Psi_{\rm P}^+\rangle_{ab}|\Phi_{\rm C}^-\rangle_{AB}}{N^-_\alpha}\Bigg)|\phi\rangle_{cC}\\
\nonumber
		&+\Bigg(\frac{|\Phi_{\rm P}^+\rangle_{ab}|\Phi_{\rm C}^-\rangle_{AB}}{N^-_\alpha}+\frac{|\Psi_{\rm P}^+\rangle_{ab}|\Phi_{\rm C}^+\rangle_{AB}}{N^+_\alpha}\Bigg)\hat{Z}|\phi\rangle_{cC}\\
\nonumber
		&+\Bigg(\frac{|\Phi_{\rm P}^-\rangle_{ab}|\Psi_{\rm C}^+\rangle_{AB}}{N^+_\alpha}-\frac{|\Psi_{\rm P}^-\rangle_{ab}|\Psi_{\rm C}^-\rangle_{AB}}{N^-_\alpha}\Bigg)\hat{X}|\phi\rangle_{cC}\\
		&+\Bigg(\frac{|\Phi_{\rm P}^-\rangle_{ab}|\Psi_{\rm C}^-\rangle_{AB}}{N^-_\alpha}-\frac{|\Psi_{\rm P}^-\rangle_{ab}|\Psi_{\rm C}^+\rangle_{AB}}{N^+_\alpha}\Bigg)\hat{X}\hat{Z}|\phi\rangle_{cC}\Bigg]
\end{align}
	where $\hat{X}$ and $\hat{Z}$ are the Pauli operators in terms of the logical qubit basis,
	$|\Psi_{\rm P}^\pm\rangle=(|HV\rangle\pm |VH\rangle)/\sqrt{2}$ and
	$|\Phi_{\rm P}^\pm\rangle=(|HH\rangle\pm |VV\rangle)/\sqrt{2}$ are the Bell states of photon-polarized states, and
	$|\Psi^\pm_{\rm C}\rangle=N^\pm_\alpha(|\alpha\rangle|-\alpha\rangle\pm|-\alpha\rangle|\alpha\rangle)$ and $|\Phi^\pm_{\rm C}\rangle=N^\pm_\alpha(|\alpha\rangle|\alpha\rangle\pm|-\alpha\rangle|-\alpha\rangle)$ with $N^\pm_\alpha=1/\sqrt{2\pm 2e^{-4|\alpha|^2}}$ are the Bell states of coherent states.
	The Bell-state measurement for the photon-polarized states, i.e., $\rm B_{s}$ for modes $a$ and $b$ in Fig.~\ref{scheme}, and another Bell-state measurement for the coherent states, i.e., $\rm B_\alpha$ for modes $A$ and $B$ in Fig.~\ref{scheme}, are performed. 
	We assume that available resources are linear optics elements with photodetectors.
	The success probability of $\rm B_{s}$ is then limited to 1/2 \cite{Calsa2001,Lutkenhaus99} while the success probability of the $\rm B_\alpha$ is $1-\exp(-2|\alpha|^2)$ \cite{Lee13}.
The process will be successful unless both the Bell-state measurements fail so that the success probability of the  teleportation of a hybrid qubit is $P_{h}=1-\exp(-2|\alpha|^2)/2$ \cite{Lee13}. To complete the teleportation process, an appropriate Pauli operation ($\mathbb{1}$, $\hat{X}$, $\hat{Z}$, or $\hat{X}\hat{Z}$) should be applied according to the measurement result
as explained in Sec.~\ref{sec:hybridqubits} ($\rm U_s$ and $\rm U_\alpha$ in Fig.~\ref{scheme}).

The time evolution of density operator $\rho$ under photon losses is governed by the Born-Markov master equation \cite{Louisell73},
	\begin{align} \label{mastereq}
		\frac{\partial\rho}{\partial\tau}=\hat{J}\rho +\hat{L}\rho,
	\end{align}
where $\tau$ is the interaction time, $\hat{J}\rho=\gamma\Sigma_{i}a_{i}\rho a_{i}^{\dagger}$, $\hat{L}\rho=-(\gamma/2)\Sigma_{i}(a_{i}^{\dagger} a_{i}\rho+\rho a_{i}^{\dagger} a_{i})$, $\gamma$ is the decay constant, and $a_{i}$ ($a_{i}^{\dagger}$) is the annihilation (creation) operator for mode $i$. The general solution of Eq.~\eqref{mastereq} is written as, $\rho(\tau)=\exp[(\hat{J}+\hat{L})\tau]\rho(0)$, where $\rho(0)$ is the initial density operator \cite{Phoenix90}.
We assume that each mode of  the channel state $|\Psi_{ch}\rangle$ suffers the same decoherence rate characterized by $\gamma$.
 The entangled channel at time $\tau$ under the above assumption is obtained using Eq.~\eqref{mastereq} as
\begin{eqnarray}
\nonumber
\rho^{ch}_{\rm I}(\tau)&=&\frac{1}{2}\big[\{\big(t^2|+\rangle\langle+|+r^2|0\rangle\langle0|\big)\otimes|t\alpha\rangle\langle t\alpha|\}^{\otimes2}\\
\nonumber&+&\{t^2e^{-2|\alpha|^2r^2}|+\rangle\langle-|\otimes|t\alpha\rangle\langle -t\alpha|\}^{\otimes2}\\
\nonumber&+&\{t^2e^{-2|\alpha|^2r^2}|-\rangle\langle+|\otimes|-t\alpha\rangle\langle t\alpha|\}^{\otimes2}\\
&+&\{\big(t^2|-\rangle\langle-|+r^2|0\rangle\langle0|\big)\otimes|-t\alpha\rangle\langle -t\alpha|\}^{\otimes2}],
\end{eqnarray}
	where $|\pm\rangle=(|H\rangle\pm|V\rangle)/\sqrt{2}$, $t=e^{-\gamma\tau /2}$, $r=\sqrt{1-e^{-\gamma\tau}}$, and $\{\cdot\}^{\otimes2}$ means the direct product of same states. As we see, coherent-state qubits not only lose their relative phase information but also undergo amplitude damping by photon losses. However, we know the value of the interaction time $\tau$, we can use $|\pm t\alpha\rangle$ as a dynamic qubit basis in order to reflect the amplitude damping as suggested in Ref.~\cite{JeongPRA2001}. Adopting this, we define a dynamic orthonormal basis of optical hybrid qubits as 
	\begin{align}	
		\{|0_L(\tau)\rangle=|+\rangle|t\alpha\rangle, |1_L(\tau)\rangle=|-\rangle|-t\alpha\rangle\},
	\end{align}
	and an unknown qubit which Alice wants to teleport as $|\phi(\tau)\rangle=\mu|0_L(\tau)\rangle+\nu|1_L(\tau)\rangle$ where $\mu=\cos(u/2)$ and $\nu=e^{iv}\sin(u/2)$. The Bell-state measurement is then performed on the input state $|\phi(\tau)\rangle$ and one part of the decohered channel state $\rho_{ch}(\tau)$. The Bell-state measurement for single photon qubits, $\rm B_s$, and that of coherent state qubits, $\rm B_\alpha$, with damped amplitudes are performed~\cite{Lee13}.  The Bell states of coherent states with damped amplitudes are 
	\begin{align}	
&	|\Psi^\pm_{\rm C}(\tau)\rangle=N^\pm_\alpha(\tau)(|t\alpha\rangle|-t\alpha\rangle\pm|-t\alpha\rangle|t\alpha\rangle),\\
& |\Phi^\pm_{\rm C}(\tau)\rangle=N^\pm_\alpha(\tau)(|t\alpha\rangle|t\alpha\rangle\pm|-t\alpha\rangle|-t\alpha\rangle),
	\end{align}
  where $N^\pm_\alpha(\tau)=1/\sqrt{2\pm 2e^{-4t^2|\alpha|^2}}$ in terms of the dynamic qubit basis. 
In order to perform $\rm B_\alpha$, a 50:50 beam splitter and two photon number resolving detectors are needed. 
We define  the 50:50 beam-splitter operator as
	\begin{equation}
		U_{i,j}=e^{-\frac{\pi}{4}(a_i^\dagger a_j-a_i a_j^\dagger )},
	\end{equation}
	where $i$ and $j$ are two field modes entering the beam splitter.
The operation of $U_{A,B}$ on coherent states is characterized as $U_{A,B}|\alpha\rangle_A|\beta\rangle_B=|(\alpha+\beta)/\sqrt{2}\rangle_A|(-\alpha+\beta)/\sqrt{2}\rangle_B$. 
	The coherent-state Bell-state measurement,  $\rm B_\alpha$, is
represented by the projection operators:
	\begin{align}
		O_{1}&=\sum\limits_{n=1}^\infty|2n\rangle_A\langle 2n|\otimes|0\rangle_B\langle 0|,\\
		O_{2}&=\sum\limits_{n=1}^\infty|2n-1\rangle_A\langle 2n-1|\otimes|0\rangle_B\langle 0|,\\
		O_{3}&=\sum\limits_{n=1}^\infty|0\rangle_A\langle 0|\otimes|2n\rangle_B\langle 2n|,\\
		O_{4}&=\sum\limits_{n=1}^\infty|0\rangle_A\langle 0|\otimes|2n-1\rangle_B\langle 2n-1|,\\
		O_{\rm e}&=|0\rangle_A\langle 0|\otimes|0\rangle_B\langle 0|,
	\end{align}
where subscripts 1, 2, 3 and 4 correspond to $\Phi^+_{\rm C}$, $\Phi^-_{\rm C}$, $\Psi^+_{\rm C}$ and $\Psi^-_{\rm C}$, respectively, while $O_{\rm e}$ represents the measurement failure for which both the detectors do not register any photon.

	The Bell-state measurement of photon-polarized states, $\rm B_s$, can be performed using the 50:50 beam splitter $U_{a,b}$ which is applied to single photon qubits of modes $a$ and $b$, two polarizing beam splitters, and four photodetectors \cite{Lutkenhaus99}. The measurement results are represented by following projective operators,
	\begin{align}
		M_{1}=&|HV\rangle_a\langle HV|\otimes|0\rangle_b\langle 0|+|0\rangle_a\langle 0|\otimes|HV\rangle_b\langle HV|, \\
		M_{2}=&|H\rangle_a\langle H|\otimes|V\rangle_b\langle V|+|V\rangle_a\langle V|\otimes|H\rangle_b\langle H|, \\
		M_{\rm e}=&|HH\rangle_a\langle HH|\otimes|0\rangle_b\langle 0|+|0\rangle_a\langle 0|\otimes|HH\rangle_b\langle HH| \nonumber \\
		&+|VV\rangle_a\langle VV|\otimes|0\rangle_b\langle 0|+|0\rangle_a\langle 0|\otimes|VV\rangle_b\langle VV|,
	\end{align}
	where $M_{1}$ and $M_{2}$ correspond to  $\Psi^+_{\rm P}$ and $\Psi^-_{\rm P}$, respectively, 
	while $M_{\rm e}$ represents a measurement failure for which both the detectors are silent. The teleportation process will be successful unless both  $\rm B_\alpha$ and  $\rm B_s$  fail.

	\begin{figure*}
		\centering
		\epsfig{file=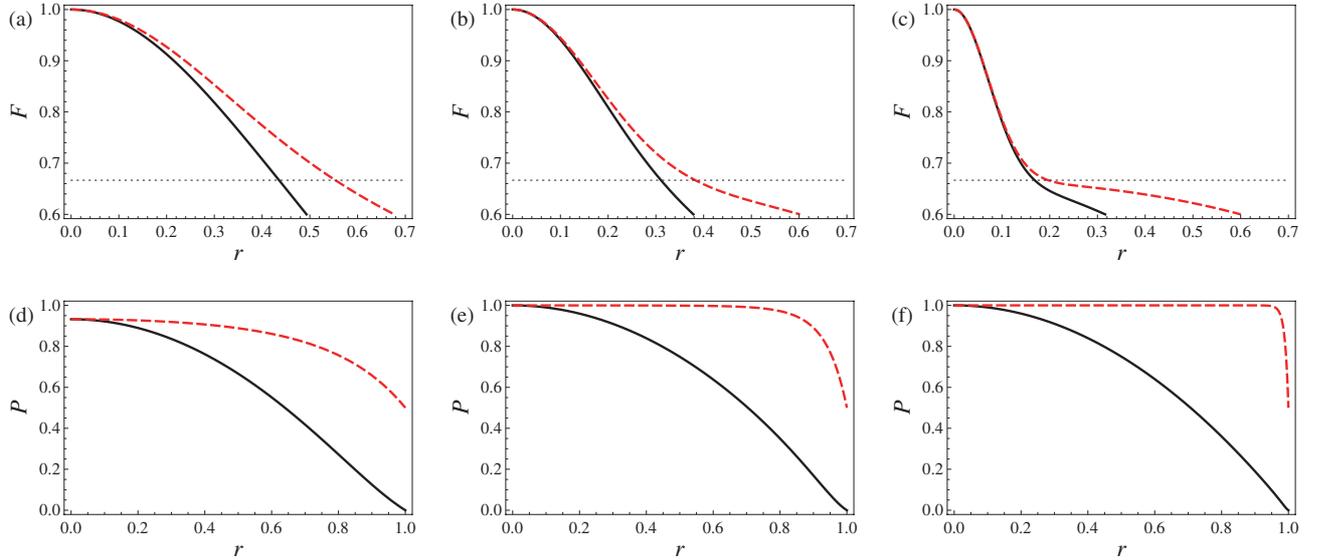,width=17cm}
		\caption{(Color online) (a)-(c) Average fidelities and (d)-(f) average success probabilities of the type-I hybrid qubits (solid curves) and the type-II hybrid qubits (dashed curves) against the normalized time $r$. The horizontal dotted lines indicate classical limit, 2/3, which can be achieved by using a separable teleportation channel. Graphs are plotted with various values of amplitude $\alpha$ of coherent states as $|\alpha|=1$ for (a) and (d), $|\alpha|=2$ for (b) and (e), and $|\alpha|=5$ for (c) and (f).}
		\label{result}
\end{figure*}
	
The unnormalized output state after measurement outcome $M_{i}\otimes O_{j}$ is obtained as
	\begin{align}
		\rho^{i,j}={\rm Tr}_{a,b,A,B}[&(U_{a,b}\otimes U_{A,B})(|\phi(\tau)\rangle\langle\phi(\tau)|\otimes\rho^{ch}_{\rm I}(\tau)) \nonumber \\
		&\times(U^\dagger_{a,b}\otimes U^\dagger_{A,B})(M_{i}\otimes O_{j})],
	\end{align}
	where the partial trace is taken over Alice's modes $a$, $b$, $A$ and $B$ in Fig.~\ref{scheme}.

   Finally, Bob should perform appropriate unitary operations ($\mathbb{1}$, $Z$, $X$, or $XZ$) according to Alice's measurement results. The details are as follows: $\mathbb{1}$ for $\rho^{1,2}_{\rm I}$ and $\rho^{{\rm e},1}_{\rm I}$, $Z$ for $\rho^{1,1}_{\rm I}$, $\rho^{1,{\rm e}}_{\rm I}$, and $\rho^{{\rm e},2}_{\rm I}$, $X$ for $\rho^{2,4}_{\rm I}$ and $\rho^{{\rm e},3}_{\rm I}$, and $XZ$ for  $\rho^{2,3}_{\rm I}$, $\rho^{2,{\rm e}}_{\rm I}$, and $\rho^{{\rm e},4}_{\rm I}$.

The final teleported state is then
\begin{eqnarray}
\nonumber \rho^T_{\rm I}(\tau)&=&|\mu|^2\big(t^2|+\rangle\langle+|+r^2|0\rangle\langle0|\big)\otimes|t\alpha\rangle\langle t\alpha|\\
\nonumber &+&t^2e^{-4|\alpha|^2r^2}\{\mu\nu^*|+\rangle\langle-|\otimes|t\alpha\rangle\langle -t\alpha|\\ \nonumber
&+&\mu^*\nu |-\rangle\langle+|\otimes|-t\alpha\rangle\langle t\alpha|\}\\
&+&|\nu|^2\big(t^2|-\rangle\langle-|+r^2|0\rangle\langle0|\big)\otimes|-t\alpha\rangle\langle -t\alpha|,
\end{eqnarray}
regardless of the outcomes of the Bell-state measurements.
The success probability is
	\begin{align}	\label{prob_s}
			P_{\rm I}(\tau)=&\sum\limits_{j=1}^2{\rm Tr}_{c,C}[\rho^{1,j}]+\sum\limits_{j=3}^4{\rm Tr}_{c,C}[\rho^{2,j}] \nonumber\\
			+&\sum\limits_{i=1}^2{\rm Tr}_{c,C}[\rho^{i,{\rm e}}]+\sum\limits_{j=1}^4{\rm Tr}_{c,C}[\rho^{{\rm e},j}] 
			=t^2(1-\frac{1}{2}e^{-2|\alpha|^2t^2}),
	\end{align} 
	where the trace was taken over for Bob's part. 
		The average fidelity between the input state $|\phi(\tau)\rangle$ and the teleported state $\rho^T_{\rm I}(\tau)$ is
\begin{eqnarray}	\label{fidelity}
	\nonumber
		F_{\rm I}(\tau)&=&\frac{1}{4\pi}\int^{2\pi}_{0}\int^{\pi}_{0}\langle\phi(\tau)|\rho^T_{\rm I}(\tau)|\phi(\tau)\rangle\sin u du dv	\\
	&=&\frac{1}{3}t^2(2+e^{-4|\alpha|^2r^2}).
\end{eqnarray}
	The results of the average fidelity and success probability are plotted in Fig.~\ref{result}.

\subsection{Teleportation of type-II hybrid qubits}

In order to perform teleportation for hybrid qubits of vacuum and single-photon states with coherent states (type-II),
the process described in Fig.~\ref{scheme} is applied again.
The total product state of an unknown input state $|\phi\rangle$ and the channel state $|\Psi_{ch}\rangle$ can be written as Eq.~(\ref{eq:prod}) by replacing $|\Psi^\pm_{\rm P}\rangle$ and $|\Phi^\pm_{\rm P}\rangle$ with $|\Psi_{\rm V}^\pm\rangle=(|01\rangle\pm |10\rangle)/\sqrt{2}$ and $|\Phi_{\rm V}^\pm\rangle=(|00\rangle\pm |11\rangle)/\sqrt{2}$, respectively. 
As discussed in Sec.~\ref{sec:hybridqubits}, we assume that when the Pauli Z operation is necessary to complete the teleportation process, we do not apply it directly on the output but rather logically relabel $|0\rangle$ and $|1\rangle$. Thus, the success probability of teleportation for type-II hybrid qubits without photon loss is the same to that for type-I qubits.

We consider quantum teleportation of type-II hybrid qubits over a lossy environment. We assume that each mode of  the channel state $|\Psi_{ch}\rangle$ suffers the same decoherence rate $\gamma$ as before. The entangled channel at time $\tau$ under above assumption is obtained using Eq.~\eqref{mastereq} as
\begin{align}
\nonumber
		\rho^{ch}_{\rm II}(\tau)&=\frac{1}{2}\big[\{\rho_{++}\otimes |t\alpha\rangle\langle t\alpha|\}^{\otimes2}
		+\{e^{-2|\alpha|^2r^2}\rho_{+-}\otimes|t\alpha\rangle\langle -t\alpha|\}^{\otimes2}\\
		&+\{e^{-2|\alpha|^2r^2}\rho_{-+}\otimes|-t\alpha\rangle\langle t\alpha|\}^{\otimes2}+\{\rho_{--}\otimes |-t\alpha\rangle\langle -t\alpha|\}^{\otimes2}\big],
\end{align}
where
\begin{align}
	\rho_{++}=&\frac{1+t}{2}|+\rangle\langle+|+\frac{r^2}{2}|+\rangle\langle-|+\frac{r^2}{2}|-\rangle\langle+|+\frac{1-t}{2}|-\rangle\langle-|,\\
	\rho_{+-}=&\frac{t^2+t}{2}|+\rangle\langle-|+\frac{t^2-t}{2}|-\rangle\langle+|,\\
	\rho_{-+}=&(\rho_{+-})^\dagger,\\
	\rho_{--}=&\frac{1-t}{2}|+\rangle\langle+|+\frac{r^2}{2}|+\rangle\langle-|+\frac{r^2}{2}|-\rangle\langle+|+\frac{1+t}{2}|-\rangle\langle-|.
\end{align}
	As before, we use the dynamic qubit basis for coherent states and define the new orthonormal basis as
	\begin{align}	
		\{|0_L(\tau)\rangle=|+\rangle|t\alpha\rangle, |1_L(\tau)\rangle=|-\rangle|-t\alpha\rangle\},
	\end{align}
	and an unknown qubit which Alice wants to teleport as $|\phi(\tau)\rangle=\mu|0_L(\tau)\rangle+\nu|1_L(\tau)\rangle$. 
	The Bell-state measurement  $\rm B_\alpha$ is performed as explained in Sec.~\ref{sec:teleportation1}.
However, in the case of type-II hybrid qubits, $\rm B_s$ is a Bell-state measurement of vacuum and single-photon states that
identifies four Bell states: $|\Psi_{\rm V}^\pm\rangle=(|01\rangle\pm |10\rangle)/\sqrt{2}$ and $|\Phi_{\rm V}^\pm\rangle=(|00\rangle\pm |11\rangle)/\sqrt{2}$.
This can be done using the 50:50 beam splitter $U_{a,b}$ and two single-photon detectors. Here, the single-photon detectors should be able to discriminate between zero, one and more than one photons.
The measurement results are represented by following projective operators,
	\begin{align}
		E_{1}&=|0\rangle_a\langle 0|\otimes|1\rangle_b\langle 1|, \\
		E_{2}&=|1\rangle_a\langle 1|\otimes|0\rangle_b\langle 0|, \\
		E_{\rm e}&=|0\rangle_a\langle 0|\otimes|0\rangle_b\langle 0|+|0\rangle_a\langle 0|\otimes|2\rangle_b\langle 2|+|2\rangle_a\langle 2|\otimes|0\rangle_b\langle 0|,
	\end{align}
	where $\Psi^+_{\rm V}$ and $\Psi^-_{\rm V}$ correspond to $E_{1}$ and $E_{2}$, and $E_{\rm e}$ represents a measurement failure. The teleportation process will be successful unless both the Bell-state measurements, $\rm B_\alpha$ and $\rm B_s$, fail.
	
	The unnormalized state after measurement outcome $E_{i}\otimes O_{j}$ is obtained as
	\begin{align}
		\rho^{i,j}_{\rm II}={\rm Tr}_{a,b,A,B}[&(U_{a,b}\otimes U_{A,B})(|\phi(\tau)\rangle\langle\phi(\tau)|\otimes\rho^{ch}_{\rm II}(\tau)) \nonumber \\
		&\times(U^\dagger_{a.b}\otimes U^\dagger_{A,B})(E_{i}\otimes O_{j})].
	\end{align}
	Bob should perform appropriate logical gate operations ($\mathbb{1}$, $Z$, $X$, or $XZ$) according to measurement results of Alice. The details are as follows: $\mathbb{1}$ for $\rho^{1,2}_{\rm II}$, $\rho^{2,1}_{\rm II}$, and $\rho^{e,1}_{\rm II}$, $Z$ for $\rho^{1,1}_{\rm II}$, $\rho^{2,2}_{\rm II}$, $\rho^{e,2}_{\rm II}$, and $\rho^{1,{\rm e}}_{\rm II}$, $X$ for $\rho^{1,3}_{\rm II}$, $\rho^{2,4}_{\rm II}$, and $\rho^{e,3}_{\rm II}$, and $XZ$ for $\rho^{1,4}_{\rm II}$, $\rho^{2,3}_{\rm II}$, $\rho^{e,4}_{\rm II}$, and $\rho^{2,{\rm e}}_{\rm II}$.
		
	The final teleported states after applying appropriate unitary transforms are different from each other according to the Bell-state measurement results. We present all possible teleported states ($\rho_{i}^T$), their probabilities ($p_i$) of obtaining such particular outcomes, and fidelities ($f_i$) with the input state $|\phi(\tau)\rangle$ in Appendix. Here we consider the average fidelity and the average success probability as
	\begin{align}
		F_{\rm II}(\tau)=&\frac{1}{4\pi}\int^{2\pi}_{0}\int^{\pi}_{0}\frac{\sum_i p_{i}f_{i}}{\sum_i p_{i}}\sin u~ {\rm d} u{\rm d}v, \label{eq:Fii}	\\
		P_{\rm II}(\tau)=&\frac{1}{4\pi}\int^{2\pi}_{0}\int^{\pi}_{0}\sum_i p_{i}\sin u ~{\rm d}u{\rm d}v,
\label{eq:Pii}
		\end{align}
	where the summations run over 1 to 5. It is difficult to perform the integration in Eq.~(\ref{eq:Fii}) in an analytical way because of the summation in the denominator, and we obtain 
the average fidelity $F_{\rm II}(\tau)$ numerically using MATHEMATICA. The average success probability in Eq.~(\ref{eq:Pii}) is obtained as $P_{\rm II}(\tau)=1-e^{-2|\alpha|^2 t^2}/2$; as one can see, the overall factor of $t^2$ in Eq.~\eqref{prob_s} is not here. While the vacuum and single-photon states in a type-II hybrid qubit after photon loss still remain in the logical qubit space, photon-polarized states in type-I hybrid qubit evolve out of the logical qubit space due to the addition of the vacuum element under photon loss effects. Such a difference between type-I and II qubits makes the drop of the factor $t^2$.
	
	We plot the average fidelity and the average success probability in Fig.~\ref{result}. We also compare these results with the results obtained with the type-I qubits in Sec.~\ref{sec:teleportation1}. Our results in Fig.~\ref{result} clearly show that the average fidelity and the average success probability for type-II are always higher than those of type-I. Again, this can be attributed to the difference in the decoherence mechanism that  a qubit of the vacuum and single-photon states (type-II) remains in the qubit space under photon loss effects, while a single-photon qubit with photon-polarized states (type-I) gets out of the qubit space.

\section{Remarks}
\label{sec:remarks}

In this paper, we have discussed two types of hybrid qubits for quantum teleportation. One is the hybrid of polarized single-photon states with coherent states (type-I), and the other is the hybrid of the vacuum and single-photon states with coherent states (type-II). Using these two different type of hybrid qubits, we have analyzed the performance of quantum teleportation taking into account both the success probability and output fidelity under the effects of photon losses on the hybrid entangled channels. We found that both the average fidelity and the success probability of teleportation using the type-II hybrid qubits are always higher than those of the type-I hybrid qubits. The reason for this result is that a type-II hybrid qubit always, even under the effects of photon losses, remains in the logical qubit space spanned by the vacuum and single-photon states. On the other hand, the leakage from the logical qubit space possibly occurs for the type-I hybrid qubits under the photon loss effects, due to the addition of the vacuum element to the photon polarization states. This difference leads to such lower fidelity and success probability for the type-I hybrid qubits. Our results show that the type-II hybrid qubits employing vacuum and single-photon states in the single photon part may be better candidates of hybrid teleportation over a lossy environment.
Our result is consistent with the previous study of single-mode qubits \cite{hoyong14} where
the qubits of the vacuum and single photon were found to be more efficient than the polarized single-photon qubits for the direct transmission and quantum teleportation.

For future studies, it will be worth investigating the performance of two different types of hybrid qubits in the implementation of scalable  quantum computation. For this, there are additional important factors to consider such as error correction models and fault-tolerant limits under the photon losses as well as resource requirements \cite{Ralph2010}. The effects of photon losses on quantum computation using the type-I hybrid qubit were already studied in Ref.~\cite{Lee13}. In a similar way, it may be possible to investigate fault-tolerant limits for the type-II hybrid qubit under the photon loss effects and compare the results with those obtained with type-I qubits. In order to analyze and compare their performance more faithfully, it may be necessary to identify an appropriate error correction model for the type-II hybrid qubits.

\section*{Acknowledgements}

This work was supported by the National Research Foundation of Korea (NRF) grant funded by the Korea government (MSIP) (No. 2010-0018295).

\begin{widetext}
\section{Appendix}
	\label{sec:appendix}
	 In this Appendix, we present all possible teleported states, their probabilities of obtaining such particular outcomes, and fidelities with the input state $|\phi(\tau)\rangle$ for the teleportation of type-II hybrid qubits. All the listed states are the final teleported states on which appropriate unitary transforms are applied. If the measurement results are revealed as $E_{1}\otimes O_{2}$, $E_{1}\otimes O_{3}$, $E_{2}\otimes O_{1}$, and $E_{2}\otimes O_{4}$, the final teleported states are
	\begin{align}
		\rho^T_1(\tau)=|\mu|^2\rho_{++}\otimes|t\alpha\rangle\langle t\alpha|+te^{-4|\alpha|^2r^2}\{\mu\nu^*\rho_{+-}\otimes|t\alpha\rangle\langle -t\alpha|
		+\mu^*\nu \rho_{-+}\otimes|-t\alpha\rangle\langle t\alpha|\}+|\nu|^2\rho_{--}\otimes|-t\alpha\rangle\langle -t\alpha|,
	\end{align}
	with the probability
	\begin{align}
		p_1(\tau)={\rm Tr}_{c,C}[\rho^{1,2}]+{\rm Tr}_{c,C}[\rho^{1,3}]+{\rm Tr}_{c,C}[\rho^{2,1}]+{\rm Tr}_{c,C}[\rho^{2,4}]=\frac{1}{4}(1-e^{-2|\alpha|^2t^2})(1+te^{-2|\alpha|^2t^2}).
	\end{align}
	Their fidelities with the input state $|\phi(\tau)\rangle$ are calculated as
		\begin{align}
			f_1(\tau)=(|\mu|^4+|\nu|^4)\frac{1+t}{2}+2|\mu|^2|\nu|^2\bigg(\frac{1-t}{2}e^{-4|\alpha|^2t^2}+t\frac{t^2+t}{2}e^{-4|\alpha|^2r^2}\bigg)+(\mu^2{\nu^*}^2+{\mu^*}^2\nu^2)t\frac{t^2-t}{2}e^{-4|\alpha|^2}+(\mu\nu^*+\mu^*\nu)\frac{r^2}{2}e^{-2|\alpha|^2t^2}.
		\end{align}
	If the measurement results are revealed as $E_{1}\otimes O_{1}$, $E_{1}\otimes O_{4}$, $E_{2}\otimes O_{2}$, and $E_{2}\otimes O_{3}$, the final teleported states are
	\begin{align}
		\rho^T_2(\tau)=|\mu|^2\rho_{++}^\prime\otimes|t\alpha\rangle\langle t\alpha|+te^{-4|\alpha|^2r^2}\{\mu\nu^*\rho_{+-}\otimes|t\alpha\rangle\langle -t\alpha|
		+\mu^*\nu \rho_{-+}\otimes|-t\alpha\rangle\langle t\alpha|\}+|\nu|^2\rho_{--}^\prime\otimes|-t\alpha\rangle\langle -t\alpha|,
	\end{align}
	where
	\begin{align}
		\rho_{++}^\prime=&\frac{1+t}{2}|+\rangle\langle+|-\frac{r^2}{2}|+\rangle\langle-|-\frac{r^2}{2}|-\rangle\langle+|+\frac{1-t}{2}|-\rangle\langle-|,\\
		\rho_{--}^\prime=&\frac{1-t}{2}|+\rangle\langle+|-\frac{r^2}{2}|+\rangle\langle-|-\frac{r^2}{2}|-\rangle\langle+|+\frac{1+t}{2}|-\rangle\langle-|,
	\end{align}
	with the probability
	\begin{align}
		p_2(\tau)={\rm Tr}_{c,C}[\rho^{1,1}]+{\rm Tr}_{c,C}[\rho^{1,4}]+{\rm Tr}_{c,C}[\rho^{2,2}]+{\rm Tr}_{c,C}[\rho^{2,3}]=\frac{1}{4}(1-e^{-2|\alpha|^2t^2})(1-te^{-2|\alpha|^2t^2}),
	\end{align}
	and the fidelities are 
		\begin{align}
			f_2(\tau)=(|\mu|^4+|\nu|^4)\frac{1+t}{2}+2|\mu|^2|\nu|^2\bigg(\frac{1-t}{2}e^{-4|\alpha|^2t^2}+t\frac{t^2+t}{2}e^{-4|\alpha|^2r^2}\bigg)+(\mu^2{\nu^*}^2+{\mu^*}^2\nu^2)t\frac{t^2-t}{2}e^{-4|\alpha|^2}-(\mu\nu^*+\mu^*\nu)\frac{r^2}{2}e^{-2|\alpha|^2t^2}.
		\end{align}
	If the measurement results are revealed as $E_{\rm e}\otimes O_{1}$ and $E_{\rm e}\otimes O_{3}$, the final teleported states are
	\begin{align}
		\rho^T_3(\tau)=|\mu|^2\rho_{++}\otimes|t\alpha\rangle\langle t\alpha|+t^2e^{-4|\alpha|^2r^2}\{\mu\nu^*\rho_{+-}\otimes|t\alpha\rangle\langle -t\alpha|
		+\mu^*\nu \rho_{-+}\otimes|-t\alpha\rangle\langle t\alpha|\}+|\nu|^2\rho_{--}\otimes|-t\alpha\rangle\langle -t\alpha|,
	\end{align}
	with the probability
	\begin{align}
		p_3(\tau)={\rm Tr}_{c,C}[\rho^{{\rm e},1}]+{\rm Tr}_{c,C}[\rho^{{\rm e},3}]=\frac{1}{4}(1-e^{-2|\alpha|^2t^2})^2,
	\end{align}
	and the fidelities are 
	\begin{align}
		f_3(\tau)=(|\mu|^4+|\nu|^4)\frac{1+t}{2}+2|\mu|^2|\nu|^2\bigg(\frac{1-t}{2}e^{-4|\alpha|^2t^2}+t^2\frac{t^2+t}{2}e^{-4|\alpha|^2r^2}\bigg)+(\mu^2{\nu^*}^2+{\mu^*}^2\nu^2)t^2\frac{t^2-t}{2}e^{-4|\alpha|^2}+(\mu\nu^*+\mu^*\nu)\frac{r^2}{2}e^{-2|\alpha|^2t^2}.
	\end{align}
	If the measurement results are revealed as $E_{\rm e}\otimes O_{2}$ and $E_{\rm e}\otimes O_{4}$, the final teleported states are	
	\begin{align}
		\rho^T_4(\tau)=|\mu|^2\rho_{++}^\prime\otimes|t\alpha\rangle\langle t\alpha|+t^2e^{-4|\alpha|^2r^2}\{\mu\nu^*\rho_{+-}\otimes|t\alpha\rangle\langle -t\alpha|
		+\mu^*\nu \rho_{-+}\otimes|-t\alpha\rangle\langle t\alpha|\}+|\nu|^2\rho_{--}^\prime\otimes|-t\alpha\rangle\langle -t\alpha|,
	\end{align}	
	with the probability
	\begin{align}
		p_4(\tau)={\rm Tr}_{c,C}[\rho^{{\rm e},2}]+{\rm Tr}_{c,C}[\rho^{{\rm e},4}]=\frac{1}{4}(1-e^{-2|\alpha|^2t^2})(1+e^{-2|\alpha|^2t^2}),
	\end{align}
	and the fidelities are 
	\begin{align}
		f_4(\tau)=(|\mu|^4+|\nu|^4)\frac{1+t}{2}+2|\mu|^2|\nu|^2\bigg(\frac{1-t}{2}e^{-4|\alpha|^2t^2}+t^2\frac{t^2+t}{2}e^{-4|\alpha|^2r^2}\bigg)+(\mu^2{\nu^*}^2+{\mu^*}^2\nu^2)t^2\frac{t^2-t}{2}e^{-4|\alpha|^2}-(\mu\nu^*+\mu^*\nu)\frac{r^2}{2}e^{-2|\alpha|^2t^2}.
	\end{align}
	Lastly, for the measurement results of $E_{1}\otimes O_{\rm e}$ and $E_{2}\otimes O_{\rm e}$, the final teleported states are
	\begin{align}
		\rho^T_5(\tau)=&\frac{1}{1-(\mu\nu^*+\mu^*\nu)r^2}\bigg[\bigg(|\mu|^2\frac{1+t}{2}-\mu\nu^*\frac{r^2}{2}-\mu^*\nu\frac{r^2}{2}+|\nu|^2\frac{1-t}{2}\bigg)\rho_{++}^\prime\otimes|t\alpha\rangle\langle t\alpha|+e^{-4|\alpha|^2r^2}\bigg\{\bigg(\mu\nu^*\frac{t^2+t}{2}+\mu^*\nu\frac{t^2-t}{2}\bigg)\rho_{+-}\otimes|t\alpha\rangle\langle -t\alpha|	\nonumber \\
		&+\bigg(\mu\nu^*\frac{t^2-t}{2}+\mu^*\nu\frac{t^2+t}{2}\bigg)\rho_{-+}\otimes|-t\alpha\rangle\langle t\alpha|\bigg\}+\bigg(|\mu|^2\frac{1-t}{2}-\mu\nu^*\frac{r^2}{2}-\mu^*\nu\frac{r^2}{2}+|\nu|^2\frac{1+t}{2}\bigg)\rho_{--}^\prime\otimes|-t\alpha\rangle\langle -t\alpha|\bigg]
	\end{align}	
	with the probability
	\begin{align}
		p_5(\tau)={\rm Tr}_{c,C}[\rho^{1,{\rm e}}]+{\rm Tr}_{c,C}[\rho^{2,{\rm e}}]=\frac{1}{2}e^{-2|\alpha|^2t^2}\{1-(\mu\nu^*+\mu^*\nu)r^2\},
	\end{align}	
	and the fidelities are 
	\begin{align}
		f_5(\tau)=&\frac{1}{1-(\mu\nu^*+\mu^*\nu)r^2}\bigg[(|\mu|^4+|\nu|^4)\bigg\{\bigg(\frac{1+t}{2}\bigg)^2+\bigg(\frac{1-t}{2}\bigg)^2e^{-4|\alpha|^2t^2}\bigg\}+2|\mu|^2|\nu|^2\bigg\{\frac{r^2}{4}(1+e^{-2|\alpha|^2t^2})^2+\bigg(\frac{t^2+t}{2}\bigg)^2e^{-4|\alpha|^2r^2}+\bigg(\frac{t^2-t}{2}\bigg)^2e^{-4|\alpha|^2}\bigg\}	\nonumber	\\
		&+(\mu^2{\nu^*}^2+{\mu^*}^2\nu^2)\bigg\{\frac{r^4}{2}e^{-2|\alpha|^2t^2}+\bigg(\frac{t^2+t}{2}\bigg)\bigg(\frac{t^2-t}{2}\bigg)(e^{-4|\alpha|^2r^2}+e^{-4|\alpha|^2})\bigg\}-(\mu\nu^*+\mu^*\nu)\frac{r^2}{2}(1+e^{-2|\alpha|^2t^2})\bigg(\frac{1+t}{2}+\frac{1-t}{2}e^{-2|\alpha|^2t^2}\bigg)\bigg].
	\end{align}
\end{widetext}

\end{document}